\documentclass[authoryear]{elsarticle}

\usepackage{amsmath, amssymb, graphics, setspace, bm, graphicx}
\usepackage[]{natbib}

\usepackage[pdftex,pdfpagemode={UseOutlines},bookmarks,bookmarksopen,colorlinks,linkcolor={blue},citecolor={blue},urlcolor={red}]{hyperref}

\newcommand{\mathsym}[1]{{}}
\newcommand{\unicode}[1]{{}}

\newcommand{\dg}{\delta}
\newcommand{\Dg}{\Delta}

\newcommand{\tg}{\theta}

\newcommand{\lam}{\lambda}
\newcommand{\pd}{\partial}

\newcommand{\der}{{\rm d}}

\newcommand{\bT}{{\bf T}}

\newcommand{\bu}{{\bf u}}
\newcommand{\vphi}{\varphi}
\newcommand{\bmu}{\bar \mu}
\newcommand{\ucrit}{u_{\rm crit}}

\newcommand{\vr}{\bm {\hat r}}

\newcommand{\bI}{{\bf I}}

\newcommand{\del}{\nabla}
\newcommand{\rg}{\rho}

\newcommand{\bxi}{{\bm \xi}}
\newcommand{\bdel}{{\bm \nabla}}

\newcommand{\MEarth}{\text{M}_\oplus}
\newcommand{\REarth}{\text{R}_\oplus}
\newcommand{\MSun}{\text{M}_\odot}

\newcommand{\bcdot}{{\bm \cdot}}

\newcommand{\be}{\begin{equation}}
\newcommand{\ee}{\end{equation}}

\begin{document}

\begin{abstract}

Plate tectonics is a geophysical process currently unique to Earth, has an important role in regulating the Earth's climate, and may be better understood by identifying rocky planets outside our solar system with tectonic activity. The key criterion for whether or not plate tectonics may occur on a terrestrial planet is if the stress on a planet's lithosphere from mantle convection may overcome the lithosphere's yield stress.  Although many rocky exoplanets closely orbiting their host stars have been detected, all studies to date of plate tectonics on exoplanets have neglected tidal stresses in the planet's lithosphere.  Modeling a rocky exoplanet as a constant density, homogeneous, incompressible sphere, we show the tidal stress from the host star acting on close-in planets may become comparable to the stress on the lithosphere from mantle convection.  Tidal stress of this magnitude may aid mantle convection stress in subduction of plates, or drive the subduction of plates without the need for mantle convective stresses.  We also show that tidal stresses from planet-planet interactions are unlikely to be significant for plate tectonics, but may be strong enough to trigger Earthquakes.  Our work may imply planets orbiting close to their host stars are more likely to experience plate tectonics, with implications for exoplanetary geophysics and habitability. We produce a list of detected rocky exoplanets under the most intense stresses. Atmospheric and topographic observations may confirm our predictions in the near future.  Investigations of planets with significant tidal stress can not only lead to observable parameters linked to the presence of active plate tectonics, but may also be used as a tool to test theories on the main driving force behind tectonic activity.

\end{abstract}


\title{The Ability of Significant Tidal Stress to Initiate Plate Tectonics}

\author[CITA,Cor,Carl]{J.J.~Zanazzi}
 \ead{jzanazzi@cita.utoronto.ca}

 \author[Birm]{Amaury H.M.J.~Triaud}
 \ead{a.triaud@bham.ac.uk}
 
 \address[CITA]{Canadian Institute for Theoretical Astrophysics, University of Toronto, 60 St George Street, ON M5S 3H8, Canada}
 \address[Cor]{Cornell Center for Astrophysics, Planetary Science, Department of Astronomy, Cornell University, Ithaca, NY 14853, USA}
 \address[Carl]{Carl Sagan Institute, Cornell University, Ithaca, NY 14853, USA}
 \address[Birm]{School of Physics \& Astronomy, University of Birmingham, Edgbaston, Birmingham B15 2TT, United Kingdom}

\date{\today}
\maketitle

\section{Introduction}

Plate tectonics plays a critical role in regulating the Earth's climate.  Mineral weathering continually reduces the amount of carbon dioxide in the Earth's atmosphere, storing carbon dioxide in the crust \citep{Southam(2015)}.  When plate tectonics pushes the Earth's crust into the mantle (subduction), this carbon dioxide is re-released into the atmosphere through volcanism.  For this reason, tectonics may play a critical role in maintaining life here on Earth (\citealt{Kasting(1993),Kopparapu(2014)}, although see \citealt{Tosi(2017)}).

Because plate tectonics plays such a crucial role on the Earth's climate, whether or not terrestrial planets can sustain plate tectonics is relevant for studies of exoplanetary climates and habitability.  The criterion for determining if a terrestrial planet may undergo plate tectonics is if the stresses on a planet's lithosphere from mantle convection may overcome the yield stress of the lithosphere to initiate subduction \citep{Korenaga(2013)}.  If this condition is not met, the lithosphere behaves as a single, rigid plate on top of the convecting mantle below (the stagnant lid regime).  In our own solar system, the Earth has active plate tectonics, while Mercury, Venus, and Mars have stagnant lids \citep{Gregg(2015)}.

The discovery of Super-Earths (planets with masses $1 \; \MEarth < M \lesssim 10 \; \MEarth$; \citealt{Batalha(2011)}) led many to consider the possibility of plate tectonics on exoplanets.  
Some early studies argued plate tectonics was impossible, since the gravity on super-Earths would make the lithosphere's yield stress quite large \citep{O'NeillLenardic(2007)}.  Other early studies argued plate tectonics was inevitable, using simulations and analytic arguments to scale different properties of the rocky super-Earth with mass, and showing mantle convective stresses can easily overcome the lithosphere's yield stress.
Later studies argued the planet's mass played a near-negligible role compared to the influence of water reducing the planet's lithospheric yield stress \citep{Korenaga(2010)}.  Numerous theoretical and numerical works have followed, some of which see mobile plate-like behavior in their models, while others do not \citep{Kite(2009),ValenciaO'Connell(2009),vanSummeren(2011),Foley(2012),O'RourkeKorenaga(2012),Stamenkovic(2012),WellerLenardic(2012),NoackBreuer(2014)}.

Many transit surveys are targeting M-dwarf stars, because temperate terrestrial exoplanets are easier to detect around them due to their close ($\lesssim 0.1 \, \text{au}$) orbital semi-major axis \citep{NutzmanCharbonneau(2008),Shields(2016)}.  M dwarfs are particularly interesting since they improve their planets' odds to transit for a given incident flux (transits will happen dozens of times per year), while producing signals that can be up to two orders of magnitude stronger (including atmospheric signatures) compared to comparable planets orbiting single Sun-like stars \citep{Rodler(2014), Barstow(2016), Morley(2017), He(2017)}. 

From exoplanetary statistics, we can infer $\sim 50 \%$ of all stars with effective temperatures cooler than $4000^\circ \, \text{K}$ harbor terrestrial planets \citep{DressingCharbonneau(2013),Bonfils(2013),DressingCharbonneau(2015),MortonSwift(2014),He(2017)}, with $\sim 20 \%$ of these planets expected to be in the habitable-zone \citep{MortonSwift(2014),DressingCharbonneau(2015)}, with some estimates pointing to higher fractions \citep{Bonfils(2013),He(2017)}.  The temperate TRAPPIST-1 planetary system \citep{Gillon(2016),Gillon(2017)}, as well as Proxima Centauri b \citep{Anglada-Escude(2016)} and Ross 128 b \citep{Bonfils(2017)}, have already been discovered orbiting M-dwarf stars.  Due to the influence of plate tectonics on a planet's climate, whether rocky planets (habitable or not) around an M-dwarf stars may support active plate tectonics is of great interest.

Super-Earths and sub-Neptune mass planets are frequent: planets more massive than Earth are encountered revolving around roughly half of Sun-like stars \citep{Mayor(2011),Howard(2012)}, and 70-80\% of M dwarfs \citep{Bonfils(2013)}, orbiting their host stars closer than Mercury orbits our Sun. Recent results examining the {\it Kepler} data point out which planets of this population are likely to be rocky ($R< 1.5 R_\oplus, m<4.5 M_\oplus$; \citealt{Rogers(2015),Fulton(2017),OwenWu(2017),Ginzburg(2017)}), raising the possibility of plate tectonics on these planets. A sub-population, called
hot super-Earths, are planets with periods of order a few days (e.g. \citealt{Fischer(2008), DawsonFabrycky(2010)}).  Because of their proximity to their host stars, these planets are likely tidally locked, with a large temperature contrast between the day and night sides of $\sim 1000^\circ \, \text{K}$ \citep{Leger(2011),CastanMenou(2011),Demory(2016a)}.  Since a planet's viscosity is temperature dependent, the day and night sides of these planets may have different tectonic activity, with a fully convective dayside without a rigid lithosphere, while the nightside may be able to sustain plate tectonics \citep{vanSummeren(2011)}. 

Many super Earths exist in multi-planetary system, called Systems of Tightly-packed Inner Planets, such as Kepler-11 \citep{Lissauer(2011)}, Kepler-33 \citep{Lissauer(2012)}, Kepler-32 \citep{Swift(2013)}, Kepler-80 \citep{MacDonald(2016)}, and Kepler-444 \citep{Campante(2015)}. STIPs are characterized by multiple planets orbiting their host stars at distances smaller than Mercury's semi-major axis,  and often have close encounters with other planets at distances $< 0.03 \, \text{au}$ that can excite their orbital eccentricities, or move them out of tidal synchronization \citep{VinsonHansen(2017),Delisle(2017)}.

Although some studies considered plate tectonics on close-in, tidally locked exoplanets \citep{vanSummeren(2011)}, the ability of tides to drive mantle convection \citep{Barnes(2009),Papaloizou(2017)}, as well as the influence of tides on aquatic life \citep{Sleep(2012),Balbus(2014),LingamLoeb(2017)}, all works to date have neglected the influence of tidal stress in a terrestrial planet's lithosphere.  As we will show, this may not be a good approximation.  The tidal influence of the host star on slightly eccentric planets with periods of order a day, as well as planets rotating non-synchronously, will generate stresses on exoplanetary lithospheres comparable to stresses from mantle convection. Tidal stresses of this magnitude may effectively weaken the planet's lithosphere aiding mantle convective stresses in subducting plates, or drive the subduction of plates without the need for mantle convective stresses, and help initiate plate tectonics on exoplanets.  We will also consider the tidal stress acting on a planet from a close passage by another planet, which may also aid tectonic activity in terrestrial exoplanets.  Section~\ref{sec:Model} sets up the model which we will use to calculate tidal stresses acting on a rocky exoplanet.  Section~\ref{sec:Subduction} explicitly calculates the forces exerted on a lithospheric plate by mantle convection and tidal stress, and discusses under what conditions these forces may overcome frictional forces and subduct plates.  Section~\ref{sec:Applications} applies this model to non-synchronously rotating exoplanets,  slightly eccentric ultra-short period planets, and systems of tightly packed inner planets.  Section~\ref{sec:ThryUncertain} discusses theoretical uncertainties in our analysis, as well as implications of our work.  In Section~\ref{sec:obs} we briefly describe various observing methods that may confirm our predictions, as well a pathway to investigate plate tectonics in regimes where tidal stresses are less relevant. Section~\ref{sec:Conc} summarizes our results.

\section{Model of Tidally Stressed Elastic Exoplanet}
\label{sec:Model}

We model the exoplanet as a constant density, incompressible sphere of mass $m$, radius $R$, and density $\rg$.  The Maxwell stress tensor, encapsulating the internal stresses of the planet, is given by \citep{LandauLifshitz(1959)}:
\be
\bT = p \bI + 2\mu \bu.
\ee
Here $p$ is the pressure, $\bI$ the identity matrix, $\mu$ is the planet's shear modulus\footnote{The shear modulus has units of pressure, and is a material property related to how the planet responds to shear stress.  The higher the planet's shear modulus, the lower the deformation when shear stress is applied to a planet}, while $\bu$ is the (incompressible) planet's strain tensor:
\be
\bu = \frac{1}{2} \big[ (\bdel \bcdot \bxi) + (\bdel \bcdot \bxi)^T \big],
\ee
where $\bxi$ is the displacement of the body from equilibrium.  We assume the planet is homogeneous ($\mu = \text{constant}$).

The planet is tidally perturbed by a body of mass $m'$ and distance $d \gg R$, with potential
\begin{align}
\phi' &\simeq \left( \frac{r}{R} \right)^2 \sum_{m=-2}^2 \Phi_{2m}' Y_{2m}(\theta,\vphi).
\label{eq:phi'}
\end{align}
Here,
\be
\Phi_{2m}' = \frac{4\pi G m' R^2}{5 d^3} Y_{2m}^* \left( \frac{\pi}{2},0 \right),
\ee
$Y_{lm}$ are spherical harmonics (with $\theta$ and $\vphi$ polar and azimuthal angles, respectively), and $d$ is the distance of the planet from the perturber.

We take hydrostatic equilibrium to be a state without shear stress.  The equations of hydrostatic equilibrium are
\begin{align}
-&\frac{\der p}{\der r} - \rg \frac{\der \phi}{\der r} = 0
\label{eq:hydro} \\
&\frac{1}{r^2} \frac{\der}{\der r} \left( r^2 \frac{\der \phi}{\der r} \right) = 4 \pi G \rg,
\label{eq:poisson}
\end{align}
where $\phi$ is the planet's gravitational potential.  The boundary conditions are $\der \phi/\der r|_{r=0} = 0$ and $p(R) = 0$.  One may show the solutions for $\phi$ and $p$ are
\begin{align}
\phi(r) &= -\pi G \rg \left( 2 R^2 - \frac{2}{3} r^2 \right),
\label{eq:phi_bare} \\
p(r) &= \frac{2\pi}{3} G \rg^2 \big(R^2 - r^2\big).
\label{eq:p_bare}
\end{align}

The planet's state of hydrostatic equilibrium is disturbed by the tidal influence of the perturber, requiring internal stresses to resist its influence.  For an incompressible, constant density, homogeneous sphere, the perturbed equations of elastostatic equilibrium are \citep{LandauLifshitz(1959)}
\begin{align}
-\bdel \dg p + \mu \del^2 \bxi - \rg \bdel (\dg \phi + \phi') &= 0, 
\label{eq:elasto} \\
\bdel \bcdot \bxi &= 0, 
\label{eq:incomp} \\
\del^2 \dg \phi &= 0,
\label{eq:dpoisson}
\end{align}
where $\dg p$ and $\dg \phi$ are the Eulerian perturbations to the planet's pressure and gravitational potential, respectively.  Equations~\eqref{eq:elasto}-\eqref{eq:dpoisson} are solved with the boundary conditions $\bxi(0) = 0$, the perturbed gravitational potential $\dg \phi$ is continuous at $r=R$, the first radial derivative of the Lagrangian perturbation to the gravitational potential is continuous at $r = R$, or
\be
\left(\frac{\pd \dg \phi}{\pd r} + 4\pi G \rho \vr \bcdot \bxi \right)_{r = R^-} = \left( \frac{\pd \dg \phi}{\pd r} \right)_{r = R^+},
\label{eq:dDgphidr}
\ee
 and the radial traction vanishes at the surface of the planet, or
\be
\vr \bcdot \Dg \bT(R) = \Big\{-\dg p \vr + \mu \vr \bcdot \left[ (\bdel \bxi) + (\bdel \bxi)^T \right] - (\bxi \bcdot \bdel p) \vr \Big\}_{r=R} = 0.
\label{eq:BCTrac}
\ee

\begin{figure}
\includegraphics[scale=1]{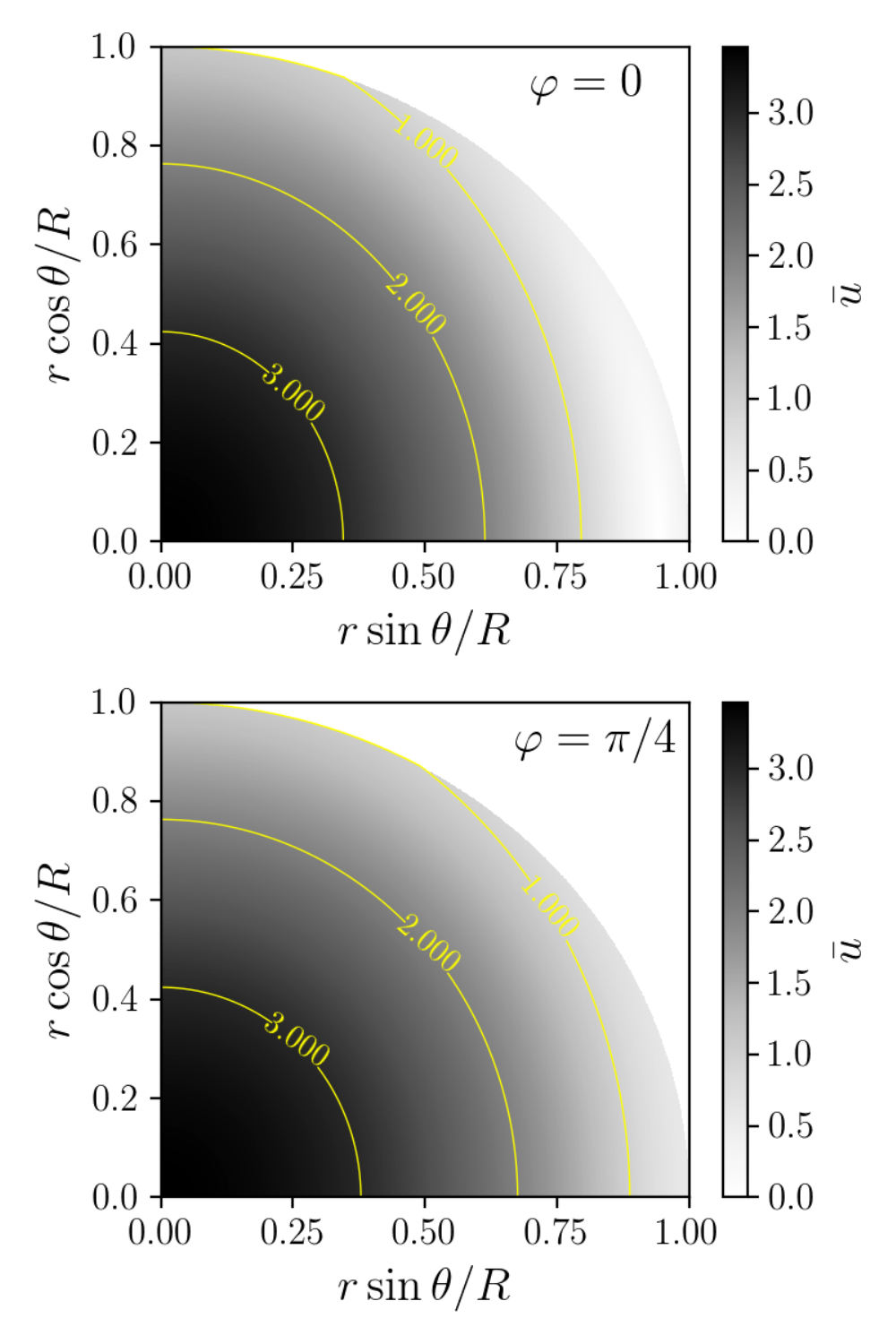}
\caption{Color map of the re-scaled strain magnitude $\bar u$ [Eq.~\eqref{eq:baru}] inside the planet, with yellow contours tracing the specific values of $\bar u$ indicated.  }
\label{fig:baru}
\end{figure}

The solution for $\bxi$ may be decomposed in the form (see \citealt{ZanazziLai(2017)})
\be
\bxi = \sum_{m=-2}^2 \left[ \xi_{r;2m} Y_{2m} \vr  + \xi_{\perp;2m} r \bdel Y_{2m} \right],
\label{eq:bxi}
\ee
where
\begin{align}
\xi_{r;2m} &= \xi_{1m} \left( \frac{r}{R} \right)^3 + 2\xi_{3m} \left( \frac{r}{R} \right) \\
\xi_{\perp;2m} &= \xi_{2m} \left( \frac{r}{R} \right)^3 + \xi_{3m} \left( \frac{r}{R} \right),
\end{align}
with $\xi_{im}$ being constants determined by the planet's boundary conditions.  In addition, we may decompose
\begin{align}
\dg \phi &= \sum_{m=-2}^2 \Phi_{2m} \left( \frac{r}{R} \right)^2 Y_{2m}(\theta,\vphi), \\
\dg p &= \sum_{m=-2}^2 P_{2m} \left( \frac{r}{R} \right)^2 Y_{2m} (\theta,\vphi),
\end{align}
with $\Phi_{2m}$ and $P_{2m}$ being undetermined constants.  Applying the boundary conditions [Eq.~\eqref{eq:dDgphidr}-\eqref{eq:BCTrac}], the solution for the constants are
\begin{align}
P_{2m} &= \frac{2 \bmu/19 - 5/2}{ \bmu + 1} \rho \Phi_{2m}', \\
\xi_{1m} &= \frac{3/2}{ \bmu + 1} \frac{\Phi_{2m}'}{g} \\
\xi_{2m} &= \frac{5/4}{ \bmu + 1} \frac{\Phi_{2m}'}{g} \\
\xi_{3m} &= - \frac{2}{ \bmu + 1} \frac{\Phi_{2m}'}{g} \\
\Phi_{2m} &= \frac{3/2}{ \bmu + 1} \Phi_{2m}',
\end{align}
where $g = Gm/R^2$, and the re-scaled rigidity $\bmu$ is given by
\begin{align}
\bmu &\equiv \frac{19 \mu}{2 \rho g R} \\
&= 2.038 \left( \frac{\mu}{10^{12} \, \text{dynes}/\text{cm}^2} \right)  \left( \frac{R}{\REarth} \right)^4 \left( \frac{m}{\MEarth} \right)^{-2}.
\end{align}
We define the dimensionless tidal bulge
\be
h \equiv \frac{G M R}{g d^3(1+\bmu)},
\label{eq:h}
\ee
and strain amplitude $u$ via
\be
u^2 \equiv \frac{1}{2} \text{Tr}(\bu \bcdot \bu) ,
\label{eq:sg}
\ee
where $\text{Tr}({\bf U})$ denotes the trace of the tensor ${\bf U}$.  In the Appendix, we explicitly calculate $u$ [Eq.~\eqref{eq:sg}] assuming $\bxi$ has the form~\eqref{eq:bxi}.  The magnitude of the strain amplitude $u(r,\theta,\vphi)$ represents how deformed the planet is under the influence of tidal stress.  Since $u \propto h$, we also define the re-scaled strain amplitude
\be
\bar u \equiv u/h.
\label{eq:baru}
\ee
In Figure~\ref{fig:baru}, we plot the re-scaled strain amplitude $\bar u$ for the elastic planet, perturbed by the tidal potential from the planet's host star.  On the surface ($r=R$), the re-scaled strain amplitude reaches a maximum of
\be
\max_{r=R,\theta,\vphi} \big[ \bar u(r,\theta,\vphi) \big] = 1.15,
\ee
while deep in the core ($r \sim 0$), the re-scaled strain amplitude reaches a maximum of
\be
\max_{r,\theta,\vphi} \big[ \bar u(r,\theta,\vphi) \big] = 3.46.
\ee
From Fig.~\ref{fig:baru}, we see the re-scaled strain $\bar u$ is of order unity almost everywhere in the planet's interior, except at $(r,\theta) \approx (R,0)$.  Since $u = h \bar u$, tidal stress causes the planet to undergo strains $u$ of order $h$ almost everywhere in the planet.

  When the planet's strain amplitude $u$ exceeds a critical value $\ucrit$ somewhere in the planet's interior, the planet can no longer maintain elasto-static equilibrium, and begins to permanently deform under the influence of the external stress acting on the planet (the von Mises yield criterion, see \citealt{TurcotteSchubert(2002)}).  
 The main driver of deformation on the Earth's lithosphere is mantle convection, which can exert shear stress exceeding $\tau_{\rm conv} \gtrsim 10^7 \, \text{dynes}/\text{cm}^2$ [see Eq.~\eqref{eq:tau_conv}], but the value of $\tau_{\rm conv}$ and exactly how it deforms the lithosphere is highly uncertain (see Sec.~\ref{sec:Subduction} for a discussion).  We take the liberal assumption that interesting tectonic activity will occur when $h \gtrsim u_{\rm crit} \sim \tau_{\rm conv}/\mu \gtrsim 10^{-5}$ (assuming $\mu \sim 10^{12} \, \text{dynes}/\text{cm}^2$).  In the next section, we explicitly calculate the forces these tidal stresses will exert on an exoplanet's lithospheric plates, and speculate on the conditions required for these forces to aid and/or hinder tectonic activity.

\section{Conditions for Initiation of Plate Tectonics}
\label{sec:Subduction}

\begin{figure}
\begin{center}
\includegraphics[scale=0.8]{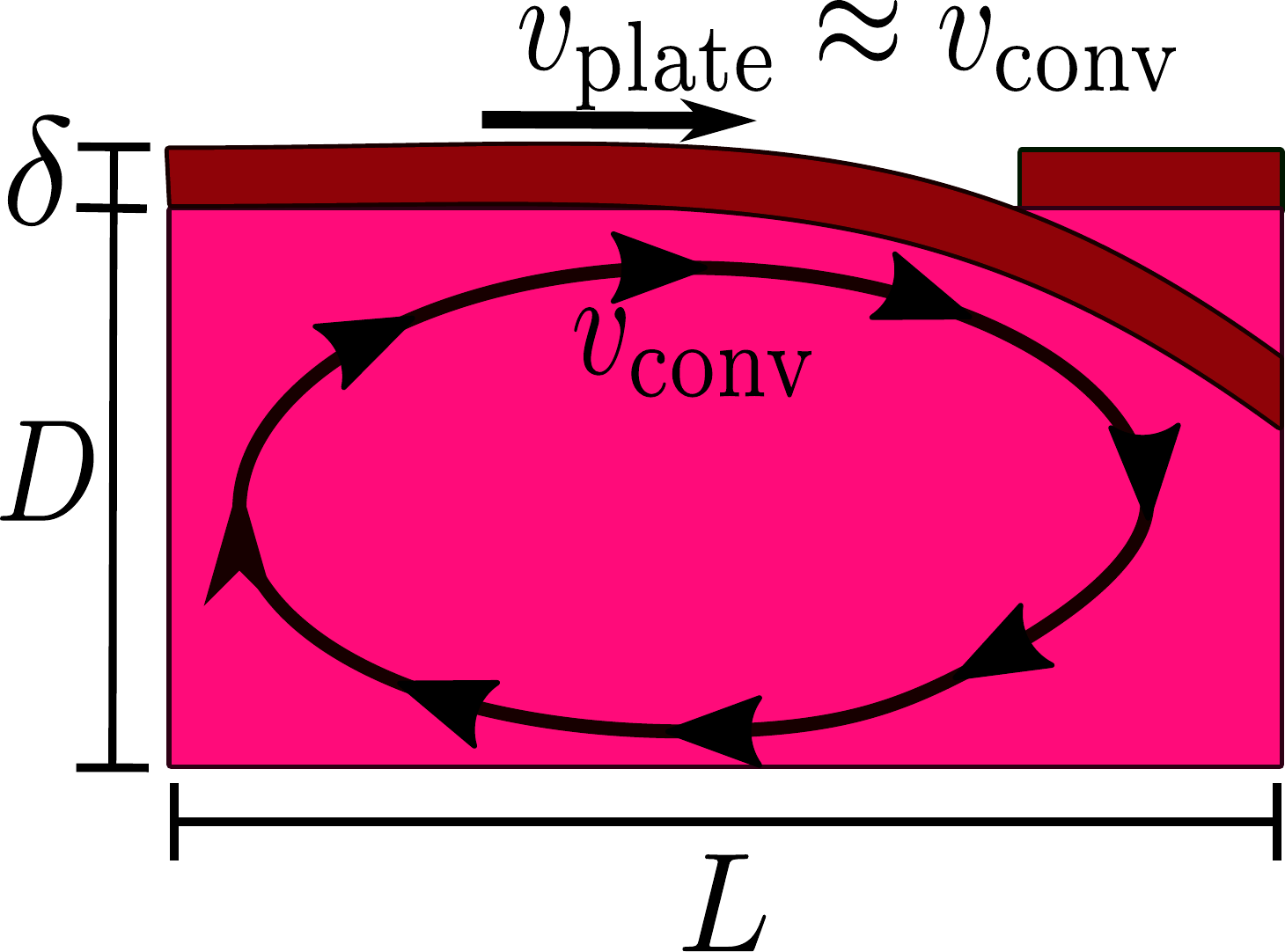}
\end{center}
\caption{
The setup for subduction of a planet's lithosphere from mantle convection.  Here, $v_{\rm conv}$ is the characteristic convective velocity in the planet's mantle, $D$ is the depth of the mantle's convective region, $L$ is the plate length, $\dg$ is the plate thickness, and $v_{\rm plate}$ is the plate velocity.
}
\label{fig:subduct}
\end{figure}

In order for a planet to undergo plate tectonics, a lateral force must push one lithospheric plate under another at a fault line (subduction).  This lateral force must overcome the frictional force acting between the two plates.  On Earth, the main lateral force is driven by stresses acting on the lithosphere from mantle convection.  This section shows under certain conditions, tidal stress may also drive subduction on extra-solar planets.

Before we discuss how tidal stress can initiate subduction, we review how stresses from mantle convection drive subduction on Earth (and potentially other planets as well).  Consider the Earth with an upper mantle of depth $D$, viscosity $\eta$, and a characteristic convective velocity $v_{\rm conv}$ under a lithospheric plate of length and width $L$, and thickness $\dg$ (see Fig.~\ref{fig:subduct} for setup).  Plates and convective cells satisfy $\dg \ll L,D \ll R$ on Earth, so we may define a local cartesian frame with ${\bm {\hat z}} = {\bm {\hat r}}$, ${\bm {\hat x}} = {\bm {\hat \theta}}$, and ${\bm {\hat y}} = {\bm {\hat \vphi}}$, and neglect how the planet's curvature affects stress and forces between plates.  We take $z = 0$ to lie on the planet's surface ($r=R$), and $x=y=0$ to lie in the center of the square plate ($-L/2 \le x,y \le L/2$).  Because the shear-stress from a fluid's viscosity $\eta$ is of order $\tau_{\rm visc} \sim \eta |\bdel {\bm v}|$ (${\bm v}$ is the fluid velocity), the shear stress on the Earth's lithosphere from mantle convection is of order
\begin{align}
&\tau_{\rm conv} \sim \eta v_{\rm conv}/D
\nonumber \\
&= 2.3 \times 10^6 \left( \frac{\eta}{10^{21} \, \text{poises}} \right) \left( \frac{v_{\rm conv}}{5 \, \text{cm}/\text{yr}} \right) \left( \frac{D}{700 \, \text{km}} \right)^{-1} \text{dynes}/\text{cm}^2.
\label{eq:tau_conv}
\end{align}
The lateral force on the lithosphere from mantle convection is then of order
\begin{align}
F_{\rm conv} = \ &\int_{-L/2}^{L/2} \int_{-L/2}^{L/2} \tau_{\rm conv} \der x \der y \sim \frac{\eta v_{\rm conv} L^2}{D}
\nonumber \\
= \ &4.4 \times 10^{22} \left( \frac{\eta}{10^{21} \, \text{poises}} \right) \left( \frac{v_{\rm conv}}{5 \, \text{cm}/\text{yr}} \right)
\nonumber \\
&\times \left( \frac{L}{1400 \, \text{km}} \right)^2 \left( \frac{D}{700 \, \text{km}} \right)^{-1} \text{dynes}.
\label{eq:Fconv}
\end{align}

The main force resisting mantle convection forces are frictional forces between two sliding plates.  The shear stress from friction acting on one plate sliding against another is \citep{Byerlee(1978),MoresiSolomatov(1998)}
\be
\tau_{\rm fric} = \tau_0 + \lam \rho g z,
\label{eq:tau_fric}
\ee
where $\tau_0$ is a cohesion parameter, and $\lam$ is the (dimensionless) friction coefficient.  Since we expect $\tau_0 \ll \lam \rho g \dg$ \citep{Byerlee(1978)}, we neglect $\tau_0$ in $\tau_{\rm fric}$ from now on.  Silicate rocks typically have $\lam \sim 0.6-0.8$ \citep{Byerlee(1978)}, but the (effective) friction coefficient may be lowered to $\lam \sim 0.03$ on Earth's surface due to hydrostatic pore pressure \citep{Korenaga(2007)}.  The frictional shear-stress exerts a lateral force between two plates of
\begin{align}
F_{\rm fric} = \ &\int_{-L/2}^{L/2} \int_{-\dg}^0 \tau_{\rm fric} \der z \der y = \frac{\lam}{2} \rho g \dg^2 L
\nonumber \\
= \ &1.2 \times 10^{24} \left( \frac{\lam}{0.03} \right) \left( \frac{\rho}{6 \, \text{g}/\text{cm}^3} \right)
\nonumber \\
&\times \left( \frac{g}{981 \, \text{cm}/\text{s}^2} \right) \left( \frac{\dg}{100 \, \text{km}} \right)^2 \left( \frac{L}{1400 \, \text{km}} \right) \text{dynes}
\label{eq:Ffric}
\end{align}
The force in~\eqref{eq:Ffric} acts in the ${\bm {\hat x}}$ direction: a force of equal magnitude acts in the ${\bm {\hat y}}$ direction as well.  Comparing Equations~\eqref{eq:Fconv} and~\eqref{eq:Ffric}, we see to initiate subduction on earth-like planets ($F_{\rm conv} \gtrsim F_{\rm fric}$), one requires $\lam$ to be weakened below the values for dry silicate rocks ($\lam \gtrsim 0.6$).

The convective stress~\eqref{eq:tau_conv} is many orders of magnitude lower than the yield stress of rocks which make up the Earth's lithosphere ($\tau_{\rm yield} \sim 10^9-10^{10} \, \text{dynes}/\text{cm}^2$), inferred through laboratory experiments \citep{Kohlstedt(1995)}.  More detailed calculations using boundary-layer theory (e.g. \citealt{TurcotteOxburgh(1967),Fowler(1993)}), as well as simulations of mantle convection (e.g. \citealt{TrompertHansen(1998),vanHeckTackley(2008),FoleyBecker(2009),WongSolomatov(2015)}) give estimates of $\tau_{\rm conv}$ to be of order $10^8 \, \text{dynes}/\text{cm}^2$.  Even with $\tau_{\rm conv}$ of this magnitude, in order to initiate subduction ($F_{\rm conv} \gtrsim F_{\rm fric}$), one requires a value of the friction coefficient $\lam$ to have a value far below that expected for lithospheric rocks ($\lam \sim 0.6-0.8$, \citealt{Byerlee(1978),Kohlstedt(1995)}).  How plate tectonics is initiated on Earth, given the stress on the Earth from mantle convection is so much lower than the lithosphere's yield stress, is an outstanding problem in Geophysics (see \citealt{Korenaga(2013)} for a review).

There are two main ideas for how to initiate plate tectonics on Earth.  One idea comes from laboratory experiments which show the presence of surface water can weaken the lithosphere's yield stress.  Experiments show minerals such as olivine which make up the Earth's lithosphere are much weaker when wet (e.g. \citealt{Karato(1986),MeiKohlstedt(2000a),MeiKohlstedt(2000b)}).  Another explanation is the simple pseudo-plastic rheologies used in simulations of plate tectonics are not realistic.  Damage theory, which models the reduction and growth of grains in a planet's lithosphere, may generate plate-like behavior with a realistic rheology \citep{BercoviciRicard(2003)}.  Both effects would work to effectively reduce friction coefficient in Equation~\eqref{eq:tau_fric}.  Due to the uncertainty of how plate tectonics is initiated and sustained here on Earth, we leave $\lam$ to be a free parameter, and note weakening mechanisms must also operate on a planet's lithosphere if the planet's friction coefficient is low ($\lam \sim 0.03$).

Tidal stress also exerts a lateral force on plates, and may potentially initiate subduction if sufficiently strong.  Because the radial traction on the planet's surface vanishes [see Eq.~\eqref{eq:BCTrac}], shear strains from tidal deformations are negligible in the planet's lithosphere ($u_{r \theta} \sim u_{\vphi r} \lesssim h \dg /R$ when $R-\dg \le r \le R$).  Therefore, the lateral force from tidal stress comes primarily from normal stresses exerting a lateral force $F_{\rm t}$ of order
\begin{align}
F_{\rm t} \sim \ &\mu h \dg L
\nonumber \\
= \ &1.4 \times 10^{24} \left( \frac{\mu}{10^{12} \, \text{dynes}/\text{cm}^2} \right) \left( \frac{h}{10^{-3}} \right)
\nonumber \\
&\times \left( \frac{\dg}{100 \, \text{km}} \right) \left( \frac{L}{1400 \, \text{km}} \right) \text{dynes}.
\label{eq:Ft}
\end{align}
Using the estimate for the tidal force acting on a lithospheric plate~\eqref{eq:Ft}, we may calculate two critical tidal bulges:
\begin{align}
h_{\rm conv} = \ &\frac{F_{\rm conv}}{\mu \dg L}
\nonumber \\
\sim \ &3.2 \times 10^{-5} \left( \frac{\eta}{10^{21} \, \text{poises}} \right) \left( \frac{v_{\rm conv}}{5 \, \text{cm}/\text{yr}} \right) \left( \frac{L}{1400 \, \text{km}} \right)
\nonumber \\
&\times \left( \frac{D}{700 \, \text{km}} \right)^{-1} \left( \frac{\mu}{10^{12} \, \text{dynes}/\text{cm}^2} \right)^{-1} \left( \frac{\dg}{100 \, \text{km}} \right)^{-1},
\label{eq:h_conv} \\
h_{\rm fric} = \ &\frac{F_{\rm fric}}{\mu \dg L}
\nonumber \\
\sim& \ 8.8 \times 10^{-4} \left( \frac{\lam}{0.03} \right) \left( \frac{\mu}{10^{12} \, \text{dynes}/\text{cm}^2} \right)^{-1}
\nonumber \\
&\times \left( \frac{\rho}{6 \, \text{g}/\text{cm}^3} \right) \left( \frac{g}{981 \, \text{cm}/\text{s}^2} \right) \left( \frac{\dg}{100 \, \text{km}} \right).
\label{eq:h_fric}
\end{align}
When $h \gtrsim h_{\rm conv}$, tidal stresses may significantly aid or hinder convective stresses in subducting plates.  When $h \gtrsim h_{\rm fric}$, tidal stresses may overcome frictional forces between two plates to initiate subduction without the aid of mantle convective stresses.

\begin{figure}
\begin{center}
\includegraphics[scale=0.8]{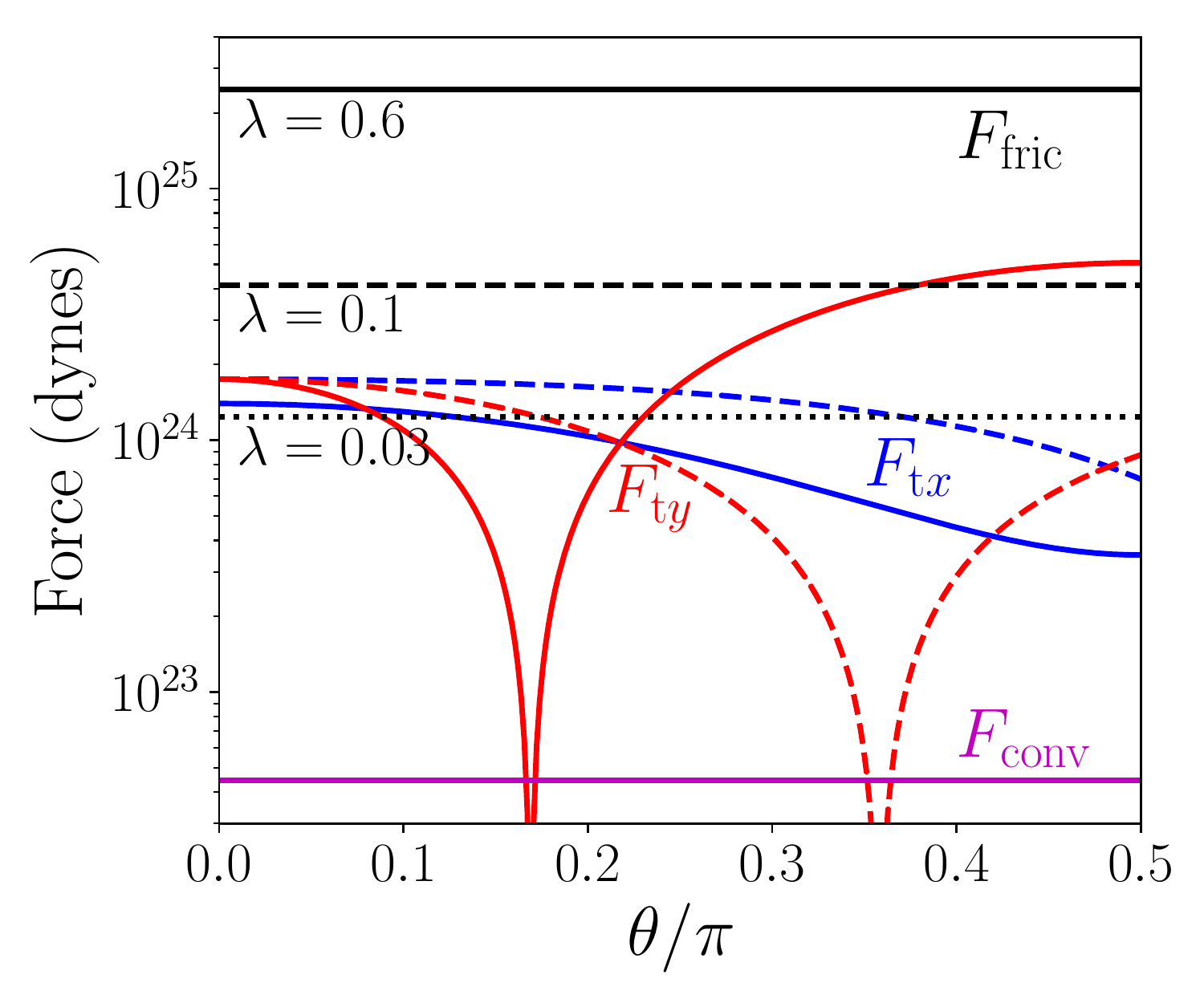}
\end{center}
\caption{
Absolute values of the tidal force components $|F_{{\rm t}x}|$ [blue, Eq.~\eqref{eq:Ftx}] and $|F_{{\rm t}y}|$ [red, Eq.~\eqref{eq:Fty}] centered at the displayed polar angles $\theta$, with azimuthal angle $\vphi=0$ (solid lines) and $\vphi=\pi/4$ (dashed lines).  Black lines denote the friction force $F_{\rm fric}$ [Eq.~\eqref{eq:Ffric}] between two sliding plates, while the magenta line denotes the mantle convection force $F_{\rm conv}$ [Eq.~\eqref{eq:Fconv}] acting on a lithospheric plate, with friction coefficients $\lam$ as indicated.  Here, $\mu = 10^{12} \, \text{dynes}/\text{cm}^2$, $h = 10^{-3}$, $\dg = 100 \, \text{km}$, $L = 1400 \, \text{km}$, $\rho = 6 \, \text{g}/\text{cm}^3$, and $g = 981 \, \text{cm}/\text{s}^2$.
}
\label{fig:TidalForces}
\end{figure}

To understand how the tidal forces acting on plates vary across an earth-like planet's surface, we explicitly calculate ${\bm F}_{\rm t} = F_{{\rm t}x} {\bm {\hat x}} + F_{{\rm t}y} {\bm {\hat y}}$ for our tidal model.  Neglecting the contribution from $u_{r\theta}$ and $u_{\theta \vphi}$ in the lateral force from tidal stress, we have
\begin{align}
F_{{\rm t}x} &= \int_{-L/2}^{L/2} \int_{-\dg}^0 \mu u_{\tg \tg}(x,y,z) \der y \der z + \int_{-L/2}^{L/2} \int_{-\dg}^0 \mu u_{\tg \vphi}(x,y,z) \der x \der z
\nonumber \\
&\simeq \mu \dg L \left[ u_{\tg \tg}(0,0,0) + u_{\tg \vphi}(0,0,0) \right],
\label{eq:Ftx} \\
F_{{\rm t}y} &= \int_{-L/2}^{L/2} \int_{-\dg}^0 \mu u_{\tg \vphi}(x,y,z) \der y \der z + \int_{-L/2}^{L/2} \int_{-\dg}^0 \mu u_{\vphi \vphi}(x,y,z) \der x \der z
\nonumber \\
&\simeq \mu \dg L \left[ u_{\tg \vphi}(0,0,0) + u_{\vphi\vphi}(0,0,0) \right],
\label{eq:Fty}
\end{align}
where we have assumed the strain $u_{ij}$ is approximately constant over the plate's volume.  Figure~\ref{fig:TidalForces} plots the magnitude of the tidal force components $|F_{{\rm t}x}|$ and $|F_{{\rm t}y}|$ of a plate centered at different locations across an earth-like planet's surface.  We see when an earth-like planet undergoes significant tidal stress ($h \gtrsim 3\times10^{-4} - 10^{-3}$), tidal forces overcome the resisting frictional force (for low $\lam$ values) and may either aid mantle convective stresses in subducting plates, or initiate subduction outright over a substantial portion of the planet's surface.

Once a plate is subducted, buoyancy forces may pull the plate into the planet's mantle \citep{TurcotteSchubert(2002)}.  Consideration of what happens to a lithospheric plate after subduction is outside the scope of this work, since an estimate of bouyancy forces would require knowledge of a planet's thermal properties.  However, we note that tidal stress will still be acting on a plate while being pulled into a planet's mantle (see Fig.~\ref{fig:baru}), potentially helping or hindering the recycling of a lithosphere.

\section{Exoplanets with High Tidal Stress}
\label{sec:Applications}

When considering how tidal stresses influence an elastic exoplanet undergoing mantle convection, three timescales need to be considered.  The first is the time it takes an elastic shear wave to propagate over the entire planet (e.g. \citealt{Quillen(2016)}):
\begin{align}
&t_{\rm elast} = \frac{R}{\sqrt{\mu/\rg}} 
\nonumber \\
&= 1.81 \times 10^{-2} \left( \frac{R}{\REarth} \right) \left( \frac{\mu}{10^{12} \, \text{dynes}/\text{cm}^2} \right)^{-1/2} \left( \frac{\rho}{6 \, \text{g}/\text{cm}^3} \right)^{1/2} \text{days},
\end{align}
We assumed in Section~\ref{sec:Model} the planet lies in elasto-static equilibrium, which requires $h^{-1} |\der h/\der t| \ll t_{\rm elast}^{-1}$.  When $h^{-1}| \der h/\der t| \gtrsim t_{\rm elast}^{-1}$, one needs to do a different calculation of the tidal stress on the planet using the impulse approximation \citep{Quillen(2016)}.

The second timescale is the time it takes a material to viscously relax to external stress (Maxwell time; \citealt{TurcotteSchubert(2002)}):
\be
t_{\rm Max} = \frac{\eta}{\mu} = 31.7 \left( \frac{\eta}{10^{21} \, \text{poises}} \right) \left( \frac{\mu}{10^{12} \, \text{dynes}/\text{cm}^2} \right)^{-1} \, \text{years},
\ee
where $\eta$ is the material's dynamic viscosity, which we have evaluated at the viscosity characteristic of the Earth's upper mantle \citep{TurcotteSchubert(2002)}.  In order for the planet to not viscously relax to the tidal stress acting on the planet, we require $h^{-1}|\der h/\der t| \gg t_{\rm Max}^{-1}$.  Viscous relaxation will limit the ability of tidal stresses to cause plastic yielding in a planet's lithosphere.

The last timescale is the turnover time of the largest convective eddies $t_{\rm conv}$:
\be
t_{\rm conv} = \frac{D}{v_{\rm conv}} = 1.4 \times 10^7 \left( \frac{5 \, \text{cm}/\text{yr}}{v_{\rm conv}} \right) \left( \frac{D}{700 \, \text{km}} \right) \text{years},
\label{eq:t_conv}
\ee
which is comparable to the timescale over which a lithospheric plate moves ($t_{\rm plate} \approx L/v_{\rm conv} \sim t_{\rm conv}$).  Tidal stress/strains may influence the large-scale convective motions in the planet's mantle when $h^{-1}|\der h/\der t| \lesssim t_{\rm conv}^{-1}$.  Since $t_{\rm conv} \gg t_{\rm Max}$ for the parameters of interest, any tidal stress which varies over the convective timescale will viscously relax before influencing convective motions in the planet.

This section directly applies our model to observed exoplanetary systems.  We emphasize the degree to which we have idealized the interior structure of these planets: we assume a constant density incompressible planet, with a uniform shear modulus which takes a value characteristic of the Earth's shear modulus.  More massive planets with rocky compositions will not be incompressible (e.g. \citealt{Seager(2007)}), and the shear modulus varies in magnitude within a body depending on the local pressure, temperature, and composition (e.g. \citealt{Dziewonski(1975),Yoder(1995),TurcotteSchubert(2002)}).  These approximations are reasonable given the level of knowledge on the interior structures and compositions of planets outside our solar system, but makes our calculation of tidal stress acting on a planet unreliable beyond an order of magnitude estimate.  For this reason, we do not explicitly state our error bars with the estimated values of tidal stresses acting on exoplanets, even though the system's parameters have well defined uncertainties.  

In the following subsections, we will consider three examples which satisfy $t_{\rm Max}^{-1} \ll h^{-1} |\der h/\der t| \ll t_{\rm elast}^{-1}$: tides from the host stars of non-synchronously rotating planets in Habitable Zones of M-dwarf's, tides from the host stars of slightly eccentric Ultra-Short Period (USP) planets, and tides from other planets in tightly packed planetary systems. The tidal stresses in these systems may aid or hinder tectonic activity when $h \gtrsim h_{\rm conv} \sim 10^{-5}$ [see Eq.~\eqref{eq:h_conv}].

\subsection{Non-synchronously rotating planets}
\label{sec:HZPlanets}

Because of their close proximities to their host stars, tides work to drive the rotation of planets toward synchronization with the planet's orbital period.  Planets may escape synchronous rotation through capture into a spin-orbit resonance \citep{GoldreichPeale(1966),Makarov(2012),Ribas(2016)}, atmospheric tides \citep{Correia(2003),Leconte(2015),Auclaire-Desrotour(2017)}, and resonant planet-planet interactions \citep{Delisle(2017),VinsonHansen(2017)}.  If this occurs, the (dimensionless) tidal bulge on the host star
\be
h = \frac{1}{1+\bmu} \left( \frac{M}{m} \right) \left( \frac{R}{a} \right)^3,
\ee
where $M$ is the stellar mass and $a$ is the planet's orbital semi-major axis, will vary on a timescale
\be
\frac{1}{h} \left| \frac{\der h}{\der t} \right| \sim \left| \frac{1}{P_{\rm orb}} - \frac{1}{P_{\rm rot}} \right|,
\ee
where $P_{\rm rot}$ is the rotation period of the planet's spin, and $P_{\rm orb}$ is the planet's orbital period.  As long as $t_{\rm Max} \gg |P_{\rm rot}^{-1} - P_{\rm orb}^{-1}|^{-1} \gg t_{\rm elast}$, the tides on the planet from the host star will generate stress of order
\be
h = \frac{1.6 \times 10^{-4}}{1+\bmu} \left( \frac{M}{0.1 \, \MSun} \right) \left( \frac{m}{1 \, \MEarth} \right)^{-1} \left( \frac{R}{\REarth} \right)^3 \left( \frac{a}{0.025 \, \text{au}} \right)^{-3}.
\label{eq:h_hab}
\ee
This tidal stress will exert a lateral force acting on a lithospheric plate, which will change in direction and magnitude as the close-in planet rotates non-synchronously

In Table~\ref{tab:hab}, we list $h$ for the temperate planets TRAPPIST-1 b, c, d, e, and f, as well as GJ 1132 b.  We see $h \gtrsim 10^{-4}-10^{-5}$ for all planets listed (assuming $\mu = 10^{12} \, \text{dynes}/\text{cm}^2$), implying tides may generate stresses of order the mantle convective stresses on these planets. We have looked through the \href{exoplanet.eu/}{Extrasolar Planets Encyclopaedia} \citep{Schneider(2011)}, and found that 42 likely rocky ($R < 1.6 \, \REarth$, \citealt{Rogers(2015),Fulton(2017)}) exoplanets have $h > 10^{-5}$, whose tectonic activity could potentially be affected by tidal processes.  We list these planets in Table~\ref{tab:known}.

\begin{table}
\centering
\begin{tabular}{ |c|c|c|c|c|c| }
\hline
Planet & $m$ ($\MEarth$) & $R$ ($\REarth$) & $P_{\rm orb}$ (days) & $h$ ($\times 10^{-5}$) \\
\hline
TRAPPIST-1 b & $ 0.86$ & $ 1.06$ & $ 1.51$ & $ 46.49 $ \\ 
TRAPPIST-1 c & $ 1.38$ & $ 1.03$ & $ 2.42$ & $ 21.11 $ \\ 
TRAPPIST-1 d & $ 0.41$ & $ 0.76$ & $ 4.05 $ & $ 4.47 $ \\ 
TRAPPIST-1 e & $ 0.64$ & $ 0.90$ & $ 6.10$ & $ 2.46 $ \\ 
TRAPPIST-1 f & $ 0.67$ & $ 1.02$ & $ 9.21$ & $ 1.08 $ \\ 
GJ 1132 b & $ 1.62$ & $ 1.13$ & $ 1.63$ & $ 50.93 $ \\ 
\hline
\end{tabular}
\caption{TRAPPIST-1  \protect\citep{Gillon(2017)} and GJ 1132 b \protect\citep{Berta-Thompson(2015)} planetary properties and dimensionless tidal bulge $h$ [Eq.~\eqref{eq:h_hab}].  TRAPPIST-1 has a mass $M = 0.08  \pm 0.01 \, \MSun $ \protect\citep{Gillon(2017)}, while GJ 1132 has a mass $M = 0.181 \pm 0.002 \, \MSun$ \protect\citep{Berta-Thompson(2015)}.  The planet's semi-major axis is computed via $a = (2\pi/P_{\rm orb})^{2/3}(GM)^{1/3}$, and we take $\mu = 10^{12} \, \text{dynes}/\text{cm}^2$ for all bodies.  We do not explicitly state uncertainties in the observational parameters and the calculated $h$ values, because of the model uncertainties (see text for discussion).}
\label{tab:hab}
\end{table}

We will discuss the prospects for detecting evidence of plate tectonics in section \ref{sec:obs}, but we will point out here that the habitability of temperate planet can be affected by tectonics, and may play an important role in stabilizing a planet's climate (e.g. \citealt{Kasting(1993), Kopparapu(2014)}). The planets TRAPPIST-1 c, d, e, and f have equilibrium temperatures between $400^\circ \, \text{K}$ and $150^\circ \, \text{K}$, considered temperate \citep{Gillon(2017)}.

\subsection{Eccentric Ultra-Short Period Planets}
\label{sec:EccUSP}

Numerous super-Earth mass planets have been detected with orbital periods $P_{\rm orb} < 1 \, \text{day}$, such as 55 Cnc e \citep{Fischer(2008),DawsonFabrycky(2010),Nelson(2014),Demory(2016b)}, CoRoT-7b \citep{Leger(2009),Bruntt(2010),Haywood(2014)}, Kepler 10b \citep{Batalha(2011),Esteves(2015),Weiss(2016a)}, WASP-47e \citep{Becker(2015),Dai(2015),Weiss(2016b),Sinukoff(2017b)}, K2 106b \citep{Sinukoff(2017a),Guenther(2017)}.  Due to their close proximities to their host stars, USP planets will be tidally circularized and rotating synchronously with rotation periods equal to their orbital periods.  This means the ``equilibrium" tidal bulge
\be
h_{\rm eq} = \frac{1}{1+\bmu} \left( \frac{M}{m} \right) \left( \frac{R}{a} \right)^3
\ee
will satisfy $\der h_{\rm eq}/\der t \simeq 0$, so $h_{\rm eq}$ will viscously relax and not stress the planet.

However, many of these USP planets have exterior planetary companions, which will pump the USP planet's eccentricity, but theoretical value for USP planetary eccentricities is uncertain.  One may delay the tidal damping of the USP planet's eccentricity by orders of magnitude through secular planet-planet interactions  \citep{WuGoldreich(2002)}.  \cite{Bolmont(2013)} argued planet-planet interactions may pump up the eccentricity of 55 Cnc e to $e \sim 10^{-3}$ - $10^{-2}$.  The combination of planet-planet interactions and exciting the USP planet's second gravitational moment ($J_2$) through tidal interactions with the host star, \cite{Rodriguez(2016)} argued the eccentricity of CoRoT-7b may be pumped up to $e \sim 0.1$.  Although the magnitude of $e$ is extremely uncertain, assuming a USP planet in a multi-planetary system has an eccentricity of order $10^{-2}$ is not unreasonable.

The small eccentricity $e$ of the USP planet will cause $h$ to vary by (assuming $e \ll 1$)
\be
\Dg h \simeq \frac{6 e}{1+\bmu} \left( \frac{m}{M} \right) \left( \frac{R}{a} \right)^3
\label{eq:Dgh_USP}
\ee
over the planet's orbital period $P_{\rm orb}$, so
\be
\frac{1}{\Dg h} \left| \frac{\der \Dg h}{\der t} \right| \sim \frac{1}{P_{\rm orb}}.
\ee
Since $t_{\rm Max} \gg P_{\rm orb} \gg t_{\rm elast}$ for an USP planet, the variation $\Dg h$ of the tidal bulge will cause stress in the planet.  Evaluating Eq.~\eqref{eq:Dgh_USP} for parameters characteristic of USP planets, we see
\be
\Dg h = \frac{3.8 \times 10^{-4}}{1+\bmu} \left( \frac{e}{0.01} \right) \left( \frac{M}{1 \, \MSun} \right) \left( \frac{m}{5 \, \MEarth} \right)^{-1} \left( \frac{R}{1.6 \, \REarth} \right)^3 \left( \frac{a}{0.015 \, \text{au}} \right)^{-3}.
\ee
Since $\Dg h \gtrsim 10^{-4}$ for a wide range of USP planet parameter space, we expect the tidal bulge variation due to the planet's small but non-zero eccentricity to cause stress comparable to the convective stresses on the planet's lithosphere.
This tidal stress will exert a lateral force on an USP planet's lithospheric plate, and the lateral force will switch direction every half orbital period.

  Table~\ref{tab:USP} lists the parameters for a number of ultra-short period (USP) planets, with the time-varying tidal bulge $\Dg h$ [Eq.~\eqref{eq:Dgh_USP}] rescaled to the orbital eccentricity $e = 0.01$.  Some of these planets have measured eccentricities, such as 55 Cnc e ($e = 0.028\substack{+0.022 \\ -0.019}$; \citealt{Nelson(2014)}), CoRoT-7b ($e = 0.12 \pm 0.07$; \citealt{Haywood(2014)}), and WASP-47e ($e = 0.03 \pm 0.02$; \citealt{Weiss(2016b)}).  But because the detection of these planetary eccentricities is marginal, we leave the planet's eccentricity fixed to a conservative value of $e = 0.01$.  We note that this exceeds the measured eccentricity of TRAPPIST-1 b ($e = 0.005 \pm 0.001$; \citealt{Luger(2017)}).

We also note that similar tidal stress should also be acting on Io, which is tidally synchronized with a small eccentricity, and displays tectonic activity (although \textit{very} different from Earth's, see \citealt{Turtle(2007)} for a review):
\begin{align}
\Dg h_{\rm Io} = \frac{4.2 \times 10^{-5}}{1+\bmu} &\left( \frac{e}{0.0041} \right) \left( \frac{M}{ {\rm M}_{\rm Jup} } \right) \left( \frac{0.015 \, \MEarth}{m} \right)
\nonumber \\
&\times \left( \frac{R}{0.286 \, \REarth} \right)^3 \left( \frac{ 0.00282 \, {\rm au}}{a} \right)^3.
\end{align}
Since $\Dg h_{\rm Io} \gtrsim 10^{-5}$, we speculate that the tidal stress from Jupiter may play a role in Io's tectonic activity.

\begin{table}
\centering
\begin{tabular}{ |c|c|c|c|c|c| }
\hline
Planet & $m$ ($\MEarth$) & $R$ ($\REarth$) & $a$ (au) & $M$ ($\MSun$) & $\Dg h \times (0.01/e)$ \\
\hline
55 Cnc e & 8.08 & 1.91 & 0.0154 & 0.95 & $2.5 \times 10^{-4}$ \\
CoRoT-7b & 4.45 & 1.58 & 0.0171 & 0.91 & $1.5 \times 10^{-4}$ \\
Kepler-10b & 3.76 & 1.47 & 0.0169 & 0.91 & $1.5 \times 10^{-4}$ \\
WASP-47e & 9.1 & 1.76 & 0.0167 & 1.00 & $1.6 \times 10^{-4}$ \\
K2-106b & 9.0 & 1.82 &0.0131 & 0.92 & $3.6 \times 10^{-4}$ \\
\hline
\end{tabular}
\caption{Ultra-short period planetary properties in multiplanet systems, with a calculated time-varying dimensionless tidal bulge $\Dg h$ [Eq.~\eqref{eq:Dgh_USP}].  We take $\mu = 10^{12} \, \text{dynes}/\text{cm}^2$ for all bodies, and rescale $\Dg h$ by $0.01/e$ due to  uncertainties in the planetary orbital eccentricities (see text for discussion).  Listed above are all Ultra-Short Period planets ($P_{\rm orb} < 1 \, \text{day}$) with detected planetary companions.  Planet and system parameters obtained from \protect\cite{Demory(2016b)} (55 Cnc e), \protect\cite{Haywood(2014)} (CoRoT-7b), \protect\cite{Esteves(2015)} (Kepler-10b), \protect\cite{Sinukoff(2017b)} (WASP-47e), and \protect\cite{Sinukoff(2017a)} (K2-106b).
}
\label{tab:USP}
\end{table}

\subsection{Tightly Packed Planetary Systems}
\label{sec:TightPlanets}

Numerous planetary systems have been discovered in which the spacing between two planetary orbits is less than $0.1$~au, such as Kepler-11 \citep{Lissauer(2011)}, Kepler-32 \citep{Swift(2013)}, Kepler-33 \citep{Lissauer(2012)}, Kepler-36 \citep{Carter(2012),Deck(2012)},  Kepler-80 \citep{MacDonald(2016)}, Kepler-444 \citep{Campante(2015)}, and TRAPPIST-1 \citep{Gillon(2016),Gillon(2017)}.  Assuming the planetary companion has a mass $m_{\rm c}$ and semi-major axis $a_{\rm c}$, with both the planet and its companion in circular orbits, the magnitude of $h$ at closest approach is
\be
h = \frac{1}{1+\bmu} \left( \frac{m_{\rm c}}{m} \right) \left( \frac{R}{|a-a_{\rm c}|} \right)^3.
\ee
Since $|a-a_{\rm c}| \ll |a+ a_{\rm c}|$, $h$ will vary by many orders of magnitude over an orbital period.  For a planetary companion with orbital period $P_{\rm c}$,
\be
\frac{1}{h} \left| \frac{\der h}{\der t} \right| \sim \left| \frac{1}{P_{\rm orb}} - \frac{1}{P_{\rm c}} \right|.
\ee
Calculating $h$ for typical tightly packed planetary system parameters, we see
\be
h = \frac{2.9 \times 10^{-6}}{1+\bmu} \left( \frac{m_{\rm c}}{m} \right) \left( \frac{R}{\REarth} \right)^3 \left( \frac{|a - a_{\rm c}|}{0.003 \, \text{au}} \right)^{-3}.
\ee

Even for very tightly packed planetary systems, $h \ll 10^{-5}$, so planet-planet tidal interactions are unlikely to affect the subduction of plates on a planet's lithosphere.  However, planet-planet tidal stresses in systems of tightly packed planets are much greater than the stress on the Earth caused by the Moon:
\be
h_{\oplus, \rm Moon} = \frac{4.8 \times 10^{-8}}{1+\bmu} \left( \frac{M}{\text{M}_{\rm Moon}} \right) \left( \frac{m}{\MEarth} \right)^{-1} \left( \frac{R}{\REarth} \right)^3 \left( \frac{a}{0.00271 \, \text{au}} \right)^{-3}.
\label{eq:h_EMoon}
\ee
If tidal stresses help induce Earthquakes here on Earth, as suggested in \cite{Ide(2016)}, planet-planet tides are likely to be a much greater source of tectonic activity on terrestrial exoplanets in systems of tightly packed planets.

\section{Theoretical Uncertainties and Implications}
\label{sec:ThryUncertain}

The biggest difference between the tidal stresses acting on short-period exoplanets and mantle convective stresses on the Earth are the timescales over which the stresses vary.  The turnover time of the convective eddies in the Earth are of order a million years [Eq.~\eqref{eq:t_conv}], and the stress from mantle convection continuously pushes a plate in one direction over this timescale.  The timescale tidal stresses on short-period exoplanets vary is less than a few years (Secs.~\ref{sec:HZPlanets}-\ref{sec:TightPlanets}), and will push a plate in alternating directions over this timescale.  Although the magnitudes of tidal forces on a short-period planet's lithosphere may be sufficient to initiate subduction, the fact that the direction these forces push a plate vary rapidly over the planet's lifetime may hinder subduction.  We speculate the rapidly varying direction of the tidal force may lead to interesting stick-slip behavior at fault lines, potentially leading to the subduction of plates.  Further work is needed to state conclusively that tidal stress aids or hinders the subduction of lithospheric plates.

We assumed the entire planet was a homogeneous, constant density incompressible elastic solid.  Although the silicate crust of USP planets may melt to form a surface lava ocean (which will have no shear modulus), these oceans will have depths $\lesssim 1\%$ the planet's radius \citep{Leger(2011)}.  Unless an USP planet has a lava ocean depth comparable to the planet's radius, the magnitude of the tidal stress calculated in Sections \ref{sec:Model} and \ref{sec:Applications} are unlikely to change by an order of magnitude.  However, we note that only the nightsides of USP planets have the right conditions for plate tectonics, since the dayside surfaces are too hot for mantle convection \citep{vanSummeren(2011)}.

We assumed a constant shear modulus and viscosity characteristic of the cold, terrestrial planets here in our solar system.  Both the shear modulus and viscosity of a planet depend on the pressure and temperature of the rocky planet (e.g. \citealt{Dziewonski(1975),Yoder(1995),TurcotteSchubert(2002)}).  Inclusion of these effects would probably require relaxing our idealized model for a rocky exoplanet and is outside the scope of this work, but will likely not change the magnitude of the tidal strain by more than an order of magnitude.

When simulating plate tectonics with a pseudo-plastic rheology, one typically needs a yield stress of order $10^8 \, \text{dynes}/\text{cm}^2$ for convective stresses to be able to subduct plates.  It is a well known problem that this is much lower than the yield strength of the rock which make up the Earth's crust, obtained from laboratory experiments \citep{Kohlstedt(1995),Korenaga(2013)}.  
 It has been argued by many surface water is the most promising way to weaken the Earth's lithospheric plates, since minerals which compose the Earth's lithosphere are weaker when wet than dry (e.g. \citealt{Karato(1986),MeiKohlstedt(2000a),MeiKohlstedt(2000b)}).  But water by itself may not be able to sufficiently weaken the Earth's lithospheric yield stress to values which can be overcome by mantle convection stresses.  As an example, the experimental work of \cite{Chen(1998)} shows wet olivine is weakened only an extra $\sim 20 \%$ compared to dry olivine in the same high pressure environments expected in the Earth's lithosphere.  This is far short of the order of magnitude drop in yield stress needed for simulations using pseudo-plastic rheologies to see plate-like behavior (e.g. \citealt{TrompertHansen(1998),vanHeckTackley(2008),FoleyBecker(2009)}).  
 This idea is further complicated by the fact that the oceanic lithosphere is expected to be dry.  Because water is highly soluble in melts, melting beneath mid-ocean ridges dehydrates the oceanic mantle and lithosphere \citep{HirthKohlstedt(1996)}.  A more promising way to create both localized weak zones and strong plates is the thermal cracking hypothesis of \cite{Korenaga(2007)}, who argues thermal stresses on a cooling early Earth are sufficient to cause deep cracks, and creating localized weak zones when the effective friction coefficient of oceanic plates is reduced due to pore fluid pressure.  The presence of surface water on exoplanets is of course a very topical question for understanding the habitability of exoplanets, and it is not clear if surface water is present or absent on the planets considered in this work.

Damage theory, which models the reduction and growth of grains in a deformed lithosphere, is a plausible way to obtain plate-like behavior with a realistic rheology, without invoking surface water to weaken the Earth's lithosphere \citep{BercoviciRicard(2003),Foley(2012)}.  Recent experimental \citep{CrossSkemer(2017)} and theoretical \citep{BercoviciSkemer(2017)} work argues the deformation of two-phase materials which make up the Earth's lithosphere reduce grain sizes, which lead to localized weak zones forming plate boundaries.  The mixing induced by damage is active in the mid lithosphere and efficient in the deep lithosphere, but is ineffective in the shallowest portions of the Earth's lithosphere \citep{BercoviciSkemer(2017)}.  We speculate that within the framework of damage theory, the fast timescales ($\lesssim \, \text{years}$) over which the tidal stress discussed in Section~\ref{sec:Applications} varies makes healing (growth of grains) almost impossible for every exoplanet of interest ($h, \Dg h \gtrsim 10^{-5}$), at least in the manner \cite{BercoviciRicard(2014)} speculated occurs on Venus because of the high surface temperature of Venus's lithosphere (explaining why Venus has plate boundaries but not active plate tectonics).  However, the fact that these tidal stresses are very un-localized (Fig.~\ref{fig:baru}) leads us to believe tidal stress by itself will not lead to the formation of sharp plate boundaries, like we have here on Earth.  Moreover, it is not clear how rapidly varying tidal stress reduces grain sizes according to the formalism developed by \cite{BercoviciRicard(2012),BercoviciRicard(2013)}.  Since mantle convection stresses vary on timescales $\gtrsim \, \text{Myr}$, the formalism developed by \cite{BercoviciRicard(2012),BercoviciRicard(2013)} assumes a purely viscous planet rheology, and neglects any mantle or lithosphere elasticity.  As we have argued in this work, the lithospheric response to varying tidal stress will be primarily elastic, not viscous.
How self-consistent modeling of damage in the lithosphere changes the results presented in Sections~\ref{sec:Model}-\ref{sec:Applications} is outside the scope of this work, but may make tectonic activity more likely for terrestrial exoplanets undergoing significant tidal stress.

 Although the exoplanetary model presented in this paper is highly idealized, one could in principal carry out a more detailed calculation of the tidal stress in a planet's lithosphere with realistic equations of state, given constraints on a planet's bulk composition and orbit.  Such calculations would illuminate the extent tidal stress effectively ``weakens'' a planet's lithosphere (or drives subduction outright), compared to planets without such significant sources of tidal stress.  If signatures of active plate tectonics were detected (see Sec.~\ref{sec:obs} for discussion), a population of tidally stressed planets could help geophysicists and planetary scientists differentiate between theories which claim weakening a planet's lithosphere with surface water or gravity is the main driver of plate tectonics \citep{O'NeillLenardic(2007),Valencia(2007),Korenaga(2010)}, or a steady internal heat source which drives mantle convection (e.g. \citealt{Barnes(2009)}). 

\section{Prospects for gathering empirical evidence of plate tectonics.}
\label{sec:obs}

The first, tentative observational evidence that geological processes may be affecting an exoplanet were produced on 55 Cnc e \citep{Demory(2016b)}. It is likely that others will follow particularly in systems that are optimal to be followed up, such as the many worlds surrounding TRAPPIST-1 \citep{Gillon(2017),deWit(2016),Barstow(2016),Morley(2017)}, or GJ\,1132b \citep{Berta-Thompson(2015), Dittmann(2017)}.

Two types of observables may be produced on exoplanets: atmospheric, and topographic. Atmospheric information can be gleaned if a planet transits its host star, by performing spectro-photometry at inferior conjunction (transmission spectroscopy), or at superior conjunction (emission spectroscopy) \citep{SeagerDeming(2010)}. In both instances, the chemical content of the atmosphere can retrieved, however different molecules sometimes have broad overlapping features that can be confused. Alternatively, for transiting and non-transiting planets, specific molecules can be identified (in transmission or emission) using high-resolution spectroscopy \citep{Snellen(2010), Rodler(2014)}. This method can uniquely identify certain molecules, but it struggles to measure relative abundances.  Light gases such as SO$_2$ are produced in large quantities by volcanic eruptions, particularly at zones of subduction, where explosive, Plinian eruptions can inject SO$_2$ high into the stratosphere where its released gases could become detectable (e.g. the Pinatubo in 1991). 

We have identified a number of planets, with high $h$ values, that are more likely than others to initiate plate tectonics thanks to significant tidal stresses (\ref{app:obs}, Table~\ref{tab:known}). We invite observers to collect atmospheric data on those, and identify common molecules indicative of plate tectonics from that subpopulation. Once tracers are identified, work can proceed on a wider range of planets in order to explore the role of mass, radius, gravity, or composition, atmospheric pressure, maybe the presence of large volumes of liquid water, on the onset and persistence of plate tectonics. Careful future studies may shed information as to why plates developed on Earth.

For a subset of planets, topographic information may also be reachable thanks to a technique called {\it eclipse mapping}. As a planet disappears behind its host star (at superior conjunction), its disc is progressively occulted by the star. High-cadence, high precision photometric time-series at ingress and egress can produce two-dimensional brightness maps of an exoplanet \citep{deWit(2012),Majeau(2012),LoudenKreidberg(2017)}.  Planet-planet occultations may reveal similar information \citep{Luger(2017)}. Active volcanoes on the surface of Io represent some of the brightest features at mid-infrared wavelengths in the Solar system \citep{PetersTurner(2013),deKleer(2014)}. Similar phenomena might become observable on planets such as TRAPPIST-1b with the JWST. Geolocating volcanoes on the surface of exoplanets, and examining their distribution may show the outlines of plates just like most volcanoes on Earth exist near subduction zones.

One potential way to infer the presence of ``tidally driven'' plate tectonics is related to the latitudinal dependence of tidal forces acting on the planet's plates.  Near the equator of the planet, the tidal force is stronger than at the north and south poles (see Fig.~\ref{fig:TidalForces}).  Numerical simulations of plate tectonics show the type of tectonic activity depends strongly on the stress and force exerted on a planet's lithosphere (e.g. \citealt{TrompertHansen(1998),vanHeckTackley(2008),FoleyBecker(2009)}).  If an exoplanet with latitude-dependent tectonic activity was detected (volcanoes tracing out plates near the equator, volcanically quiet near the poles), this may be a signature of tidally-driven plate tectonics.  However, numerical simulations are needed before any concrete predictions are made.

\section{Conclusions}
\label{sec:Conc}

What conditions are necessary for the onset and sustenance of plate tectonics is a question of interest to both geophysicists and astrophysicists looking for worlds outside our solar system.  Many planets thought to be primarily rocky orbit close ($\lesssim 0.1 \, \text{au}$) to their host stars.  By modeling an exoplanet as a constant density, homogeneous, incompressible sphere, we calculate the tidal stresses and strains acting everywhere in the planet (Sec.~\ref{sec:Model}; Fig.~\ref{fig:baru}).  
  We then show when tidal stress is sufficiently strong, tidal forces exerted on lithospheric plates can be stronger than the frictional force between two sliding plates (Sec.~\ref{sec:Subduction}; Fig.~\ref{fig:TidalForces}).  Tidal forces have the potential to aid mantle convective stresses in subducting lithospheric plates, or drive subduction without the need for mantle convective stresses, initiating plate tectonics on exoplanets.  Rocky planet rotating non-synchronously around M-dwarfs, and eccentric Ultra-Short Period ($P_{\rm orb} \lesssim 1 \, \text{day}$) planets may have interesting tectonic activity due to the significant tidal stress induced on the planet by the host star (Secs.~\ref{sec:HZPlanets}-\ref{sec:EccUSP}).
Planet-planet tides in systems of tightly packed inner-planets do not induce sufficient tidal stress to significantly aid tectonic activity.  However, we note that these stresses remain significantly larger than the tidal stress on the Earth from the Moon, potentially implying planet-planet tidal interactions at conjunction are capable of triggering Earthquakes on exoplanets (Sec.~\ref{sec:TightPlanets}).
Many uncertainties remain on the plausibility of tidal stress to drive tectonic activity, the biggest being the rapid variation in direction and magnitude of tidal forces exerted on lithospheric plates (Sec.~\ref{sec:ThryUncertain}).  We discuss prospects for gathering observational evidence for tectonic activity on exoplanets in Section~\ref{sec:obs}.
  This work has implications for the geology and habitability of terrestrial exoplanets orbiting close ($\lesssim 0.1 \, \text{au}$) to their host stars.

\section*{Acknowledgements}

We thank the anonymous referees whose comments improved the quality and clarity of this work. J. J. Zanazzi thanks Hilke Schlichting, Dong Lai, and Lauren Weiss for useful conversations.  Amaury Triaud would like to thank the entire TRAPPIST/SPECULOOS team for avid discussions.  JZ is supported in part by a NASA Earth and Space Sciences Fellowship in Astrophysics.  We thank United Airlines for their hospitality in providing a venue where a significant portion of this work was initiated, and for their seat allocation algorithm, that placed us next to each other.

\appendix

\section{Explicit Calculation of Strain Amplitude}

Decomposing the symmetric tensor $\bu$ into spherical coordinates \citep{LandauLifshitz(1959)}, we may write Eq.~\eqref{eq:sg} as
\be
u^2 = \frac{1}{2} \left( u_{rr}^2 + u_{\theta\theta}^2 + u_{\vphi\vphi}^2 \right) + u_{\theta\vphi}^2 + u_{r\theta}^2 + u_{\vphi r}^2.
\ee
Letting $i,j \in \{r,\theta,\vphi\}$, we may decompose $u_{ij}$ as
\be
u_{ij} = u_{ij;20} + 2 u_{ij;22},
\ee
where $u_{ij;lm}$ denotes a strain term proportional to the displacements $\xi_{r;lm}$ and $\xi_{\perp;lm}$ [see Eq.~\eqref{eq:bxi}].  Explicitly, we have
\begin{align}
u_{rr;20} &= A_{20} \frac{\der \xi_{r;20}}{\der r} (3\cos^2\theta - 1), \\
u_{\theta\theta;20} &= A_{20} \frac{\xi_{r;20}}{r} (3 \cos^2\theta - 1) - 6 A_{20} \frac{\xi_{\perp;20}}{r} (\cos^2 \theta - \sin^2\theta), \\
u_{\vphi\vphi;20} &= -6 A_{20} \frac{\xi_{\perp;20}}{r} \cos^2\theta + A_{20} \frac{\xi_{r;20}}{r} (3 \cos^2\theta - 1), \\
u_{\theta\vphi;20} &= u_{\vphi r;20} = 0, \\
u_{r\theta;20} &= -3 A_{20} \left( \frac{\der \xi_{\perp;20}}{\der r} - \frac{\xi_{\perp;20}}{r} + \frac{\xi_{r;20}}{r} \right) \cos\theta \sin\theta,
\end{align}
and
\begin{align}
u_{rr;22} &= A_{22} \frac{\der {\xi_{r;22}}}{\der r} \sin^2 \theta \cos 2 \vphi, \\
u_{\theta\theta;22} &= A_{22} \frac{\xi_{r;22}}{r} \sin^2 \theta \cos 2\vphi + 2 A_{22} \frac{\xi_{\perp;22}}{r} (\cos^2\theta - \sin^2\theta) \cos 2 \vphi, \\
u_{\vphi\vphi;22} &= A_{22}\frac{\xi_{r;22}}{r} \sin^2\theta \cos2\vphi - 2 A_{22} \frac{\xi_{\perp;22}}{r} (\cos^2\theta - \sin^2\theta) \cos 2\vphi, \\
u_{\theta\vphi;22} &= -2 A_{22} \frac{\xi_{\perp;22}}{r} \cos\theta \sin2\vphi, \\
u_{r\theta;22} &= A_{22} \left( \frac{\der \xi_{\perp;22}}{\der r} - \frac{\xi_{\perp;22}}{r} + \frac{\xi_{r;22}}{r} \right) \sin\theta \cos\theta \cos2\vphi, \\
u_{\vphi r;22} &= - A_{22} \left( \frac{\xi_{r;22}}{r} + \frac{\der \xi_{\perp;22}}{\der r} - \frac{\xi_{\perp;22}}{r} \right) \sin \theta \sin 2\vphi,
\end{align}
where
\be
A_{20} = \sqrt{ \frac{5}{16 \pi} },
\hspace{5mm}
A_{22} = \sqrt{ \frac{15}{32 \pi} }.
\ee

\section{Calculation of tidal stresses for known exoplanets}
\label{app:obs}

\begin{table}
\centering
\begin{tabular}{ |c|c|c|c|c|c| }
\hline
Name & $m$ ($\MEarth$) &  $R$ ($\REarth$) & $P_{\rm orb}$ (days) & $M$ ($\text{M}_\odot$) & $h$ ($\times 10^{-5}$) \\
\hline
CoRoT-7 b & $ 4.74$ & $ 1.49$ & $ 0.85$ & $ 0.93$ & $ 228.24 $ \\ 
EPIC 246393474 b & $ 5.31$ & $ 1.50$ & $ 0.28$ & $ 0.66 $ & $ 2044.65 $ \\ 
GJ 1132 b & $ 1.62 $ & $ 1.13$ & $ 1.63 $ & $ 0.181 $ & $ 50.93 $ \\ 
GJ 9827 b & $ 8.20$ & $ 1.60 $ & $ 1.21$ & $ 0.66$ & $ 98.31 $ \\ 
GJ 9827 c & $ 2.51$ & $ 1.26 $ & $ 3.65$ & $ 0.66 $ & $ 11.36 $ \\ 
HD 3167 b & $ 5.02 $ & $ 1.67 $ & $ 0.96 $ & $ 0.08 $ & $ 212.19 $ \\ 
K2-106 b & $ 8.36 $ & $ 1.49 $ & $ 0.57 $ & $ 0.93$ & $ 365.85 $ \\ 
KOI-1843 b & $ 0.32 $ & $ 0.57 $ & $ 0.18 $ & $ 0.46 $ & $ 2045.96 $ \\ 
KOI-2700 b & $ 0.86 $ & $ 1.04 $ & $ 0.91 $ & $ 0.63 $ & $ 128.56 $ \\ 
Kepler-10 b & $ 3.33 $ & $ 1.44 $ & $ 0.84 $ & $ 0.91 $ & $ 245.87 $ \\ 
Kepler-100 b & $ 7.34 $ & $ 1.28 $ & $ 6.89 $ & $ 1.11 $ & $ 1.87 $ \\ 
Kepler-102 b & $ 0.41 $ & $ 0.46 $ & $ 5.29 $ & $ 0.81 $ & $ 1.89 $ \\ 
Kepler-102 d & $ 2.61 $ & $ 1.14 $ & $ 10.31 $ & $ 0.81 $ & $ 1.22 $ \\ 
Kepler-105 c & $ 4.45 $ & $ 1.28 $ & $ 7.13 $ & $ 1.28 $ & $ 2.52 $ \\ 
Kepler-114 b & $ 7.00 $ & $ 1.16 $ & $ 5.19 $ & $ 0.71 $ & $ 2.62 $ \\ 
Kepler-20 e & $ 3.08 $ & $ 0.85 $ & $ 6.10 $ & $ 0.91 $ & $ 1.64 $ \\ 
Kepler-21 b & $ 5.09 $ & $ 1.60 $ & $ 2.79 $ & $ 1.34 $ & $ 23.63 $ \\ 
Kepler-338 e & $ 8.58 $ & $ 1.55 $ & $ 9.34 $ & $ 1.10 $ & $ 1.48 $ \\ 
Kepler-406 b & $ 6.36 $ & $ 1.40 $ & $ 2.43 $ & $ 1.07 $ & $ 21.28 $ \\ 
Kepler-406 c & $ 2.71 $ & $ 0.83 $ & $ 4.62 $ & $ 1.07 $ & $ 3.03 $ \\ 
Kepler-408 b & $ 6.36 $ & $ 0.80 $ & $ 2.57 $ & $ 1.08 $ & $ 4.14 $ \\ 
Kepler-42 b & $ 2.86 $ & $ 0.77 $ & $ 1.21 $ & $ 0.13 $ & $ 34.02 $ \\ 
Kepler-42 c & $ 1.91 $ & $ 0.71 $ & $ 0.45 $ & $ 0.13 $ & $ 278.01 $ \\ 
Kepler-42 d & $ 0.95 $ & $ 0.56 $ & $ 1.86 $ & $ 0.13 $ & $ 15.03 $ \\ 
Kepler-445 b & $ 6.36 $ & $ 1.55 $ & $ 2.98 $ & $ 0.18 $ & $ 17.45 $ \\ 
Kepler-445 d & $ 3.50 $ & $ 1.23 $ & $ 8.15 $ & $ 0.18 $ & $ 1.99 $ \\ 
Kepler-446 b & $ 4.45 $ & $ 1.47 $ & $ 1.57 $ & $ 0.22 $ & $ 67.67 $ \\ 
Kepler-446 c & $ 2.86 $ & $ 1.08 $ & $ 3.04 $ & $ 0.22 $ & $ 12.16 $ \\ 
Kepler-446 d & $ 3.18 $ & $ 1.32 $ & $ 5.15 $ & $ 0.22 $ & $ 5.80 $ \\ 
Kepler-60 b & $ 4.19 $ & $ 1.68 $ & $ 7.13 $ & $ 1.10 $ & $ 3.97 $ \\ 
Kepler-62 b & $ 8.90 $ & $ 1.28 $ & $ 5.71 $ & $ 0.69 $ & $ 2.34 $ \\ 
Kepler-70  b & $ 4.45 $ & $ 0.75 $ & $ 0.24 $ & $ 0.50 $ & $ 539.53 $ \\ 
Kepler-70 c & $ 0.67 $ & $ 0.86 $ & $ 0.34 $ & $ 0.50 $ & $ 794.95 $ \\ 
Kepler-78 b & $ 1.69 $ & $ 1.17 $ & $ 0.36 $ & $ 0.81 $ & $ 1109.18 $ \\ 
Kepler-93 b & $ 4.00 $ & $ 1.45 $ & $ 4.73 $ & $ 0.91 $ & $ 7.51 $ \\ 
LP 358-499 b & $ 2.19 $ & $ 1.28 $ & $ 3.07 $ & $ 0.52 $ & $ 16.35 $ \\ 
LP 358-499 c & $ 3.18 $ & $ 1.45 $ & $ 4.87 $ & $ 0.52 $ & $ 7.35 $ \\ 
TRAPPIST-1 b & $ 0.86 $ & $ 1.06 $ & $ 1.51 $ & $ 0.08 $ & $ 46.49 $ \\ 
TRAPPIST-1 c & $ 1.38 $ & $ 1.03 $ & $ 2.42 $ & $ 0.08 $ & $ 21.11 $ \\ 
TRAPPIST-1 d & $ 0.41 $ & $ 0.76 $ & $ 4.05 $ & $ 0.08 $ & $ 4.47 $ \\ 
TRAPPIST-1 e & $ 0.64 $ & $ 0.90 $ & $ 6.10 $ & $ 0.08 $ & $ 2.46 $ \\ 
TRAPPIST-1 f & $ 0.67 $ & $ 1.02 $ & $ 9.21 $ & $ 0.08 $ & $ 1.08 $ \\ 
\hline
\end{tabular}
\caption{Exoplanets with with $m < 10 \, \MEarth$, $R < 1.6 \, \REarth$, and $h > 10^{-5}$ [Eq.~\eqref{eq:h_hab}], gathered from the \href{exoplanet.eu/}{Extrasolar Planets Encyclopaedia} database \protect\citep{Schneider(2011)}.  Here, $m$ is the planet's mass, $R$ is the planet's radius, $P_{\rm orb}$ is the planet's orbital period, and $M$ is the host star's mass.  The planet's semi-major axis is computed via $a = (2\pi/P_{\rm orb})^{2/3}(GM)^{1/3}$, and we take $\mu = 10^{12} \, \text{dynes}/\text{cm}^2$ for all bodies.}
\label{tab:known}
\end{table}

\end{document}